\documentclass[12pt,preprint]{aastex}

\newcommand\ms{M$_{\odot}$}
\slugcomment{The Astrophysical Journal, in press}

\shorttitle{Intergalactic Stellar Population}
\shortauthors{Stanghellini et~al.}
  
\begin{document}
  
\title{Merging of Elliptical Galaxies as Possible Origin of the Intergalactic 
Stellar Population}
\author{Letizia Stanghellini}
\affil{National Optical Astronomy Observatory, 950 N. Cherry Av.,
Tucson, AZ  85719}
\email{lstanghellini@noao.edu}
\author{A. Cesar Gonz\'alez-Garc\'{\i}a}
\affil{Instituto de Astrofisica de Canarias, v\'{\i}a L\'actea s/n, 
La Laguna, E-38200 Tenerife, Spain}
\email{cesar.gonzalez@iac.es}
\and
\author{Arturo Manchado}
\affil{
Instituto de Astrof\'isica de Canarias, v\'{\i}a L\'actea s/n, 
La Laguna, E-38200 Tenerife, Spain}
\email{amt@iac.es}

\begin{abstract}
We present N-body simulations of elliptical galaxy encounters into dry mergers
to study the resulting unbound intergalactic stellar population, in particular
that of the post-Main Sequence stars. 
The systems studied are pairs of spherical galaxies without
dark halos. The stellar content of the model galaxies
is distributed into mass-bins representing low- and intermediate-mass 
stars (0.85 -- 8 \ms) according to Salpeter's initial mass function. 
Our models follow the dynamical evolution of galaxy encounters colliding 
head-on from initial low-energy parabolic or high-energy mildly-hyperbolic orbits,
and for a choice of initial-mass ratios.
The merging models with initial parabolic orbits have $M_2/M_1$=1 and 10, 
and they leave behind respectively 5.5$\%$ and 10$\%$ of the total initial mass as unbound stellar 
mass.
The merging model with initial hyperbolic orbit has $M_2/M_1$=1, and 
leaves behind 21$\%$ of its initial stellar mass as unbound mass, 
showing that the efficiency in producing 
intergalactic stars through a high-energy hyperbolic 
encounter is about four times than through a parabolic encounter of the
same initial mass ratio.
By assuming that all progenitor galaxies as well as the merger remnants 
are homologous systems we obtained that the 
intergalactic starlight is 17$\%$ and 28$\%$ of the total starlight
respectively for the parabolic and hyperbolic encounters 
with $M_2/M_1$=1.
In all models, different mass stars have the same 
probability of becoming unbound and feeding the intergalactic stellar
population.

\end{abstract}

\keywords{Galaxies: elliptical and lenticular, cD; interactions; stellar content.
Stars: AGB and post-AGB. Planetary nebulae: general.}

\section{Introduction}

Intergalactic (IG) starlight, both diffuse and from resolved
stars, has been intensively studied 
in the past decade in the IG medium of both groups and clusters of galaxies.
The long-lived populous class of low- and intermediate-mass stars, in the 
form of red giants (RGB and AGB), planetary nebulae (PNe), and their diffuse light, has been
observed in the IG medium of different types of galaxy associations, 
from poor groups like the M81
group of galaxies
(Feldmeier et~al. 2004a) and the Leo region corresponding to the HI cloud
(Castro-Rodr{\'{\i}}guez et~al. 2003), to compact, tidal groups 
(White et~al. 2003),
up to the Virgo (Ferguson et~al. 1998; Durrell et~al. 2002; 
Feldmeier et~al. 1998, 2003, 2004b)
and Coma (Gerhard et~al. 2005) clusters, as well as in higher redshift 
clusters (Zibetti et~al. 2005). 

The observations are telling us that a considerable fraction of post-Main Sequence (PMS)
stars in galaxy associations
is found between galaxies. The contribution of the
unbound stars to the total mass and light of the association
varies greatly, depending mainly 
on the galaxy concentration of the considered cluster or group. 
Observations of poorly populated groups seem to indicate
that the upper limit of the IG contribution to the total light 
is very low (up to 1.6$\%$, Castro-Rodr{\'{\i}}guez et~al. 2003), while
surveys of IG PNe, AGB and RGB stars in the nearby clusters indicate
that the intra-cluster (IC) starlight contribution is roughly 5 to 20$\%$ of the 
total starlight (Durrell et~al. 2002; Aguerri et~al. 2005; 
Feldmeier et~al. 2004b), 
depending
on the assumptions made on the stellar populations probed and on the completeness of the 
sample surveyed. The observed range of the fraction of 
IG starlight is also supported by the observations of diffuse starlight in clusters 
at z$\sim$0.25 (Zibetti et~al. 2005). 

A host of explanations for the origin of the IG starlight have been proposed in a 
variety of studies. The tidal interactions between galaxies have been explored in some depth
by several Authors, among whose Merritt (1983) and Moore et~al. (1996). An alternative 
scenario that is well suited for the IC environment has been proposed by
Muccione and Ciotti (2004), where the stellar stripping from galaxies is driven by
interactions between the stellar orbits within the galaxies and the cluster tidal field.
Among all possible explanations for 
the existence of the IG starlight, the scenario of the elliptical galaxy merging
({\it dry merging}) has been the least explored. 

In this paper we present numerical models of
dry merging of galaxies with different initial mass ratios, aimed at
describing possible scenarios for the production of the unbound stellar mass in the IG medium. 
Our models are crafted in particular to account for the PMS IG population
that is produced by the merging of elliptical galaxy pairs.
The red galaxy merging scenario may not be very common at the present time
in the core 
of galaxy
clusters, where the velocity dispersion are of the order of 1000 km s$^{-1}$. 
Nonetheless,
red mergers have very recently proved to be the common evolutionary
path to field (van Dokkum 2005) and cluster 
(Tran et~al. 2005)
elliptical galaxies, and may be also the path to produce IG starlight in galaxy groups, or in 
clusters periphery, where the
the velocity dispersions is typically much lower than in young galaxy clusters cores
(Arnaboldi et~al. 2004).
Dry mergers have been recently studied by  Gonz\'alez-Garc\'{\i}a \& van Albada (2005ab), 
and have proved to preserve the properties of the elliptical galaxy fundamental plane 
(Gonz\'alez-Garc\'{\i}a \& van Albada 2003; Nipoti et~al. 2003), providing
a further handle to
study mergers and stellar populations.

In Section 2 we present our models, including the type of numerical experiment performed,
the astrophysical input parameters, the considerations on stellar populations, the initial
conditions and constraints, and a description of the methodology and stability tests of 
the numerical models. Section 3 illustrates our results, with particular attention to
the IG stellar population produced. The discussion is in $\S$4, where we present a 
limited
comparison with the observational data and the likelihood that the 
dry-merging scenario might account for the observed IG starlight. This paper represent a first
attempt at modeling the IG population with dry merging of galaxies, and the
parameter space is thus exploratory. More detailed models, and a larger
array of calculated observable parameters, will be presented in forthcoming papers.

\section{Models}

\subsection{Galaxies, Initial conditions}

We perform numerical models of three cases of dry mergers, with a choice of
initial mass ratios $M_2/M_1$=1 or 10 \footnote{In this paper the suffix $_1$ always refers
to the less massive galaxy of the merging pair.}. 
The 
initial galaxy conditions are similar to those described by Gonz\'alez-Garc\'{\i}a \& van Albada 
(2005a), where a systematic study of the encounters between two spherical 
systems without a dark matter halo was performed. These models without dark matter are a good
first approximation to study the IG population in galaxy associations.
In fact, Gonz\'alez-Garc\'{\i}a \& van Albada (2005b) have shown that merging 
of elliptical galaxies
with dark halos produces a luminous particle distribution that is very similar to that 
resulting from the merging of elliptical galaxies without dark matter (Gonz\'alez-Garc\'{\i}a \& van Albada 
2005a). On the other hand, similar models of those presented here, but with the inclusion of dark matter
will be performed in the future to confirm our results, since Gonz\'alez-Garc\'{\i}a \& van Albada (2005ab)
did not compare the unbound stellar population derived from the merging processes with and
without the inclusion of dark matter.

We use isotropic spherical Jaffe (1983) models as initial conditions for our experiments, and the
algorithm developed by Smulders \& Balcells (unpublished, 
see Gonz\'alez-Garc\'{\i}a \& van Albada 2005a for a detailed description of the algorithm). 
The projected surface mass density (hereafter, surface density)
of such models presents a slope that decreases roughly as $R^{1/4}$, 
which makes it a suitable representation for elliptical galaxies, although the central parts 
present a cusp. The distribution function presents an analytical solution for the collision-less 
Boltzmann equation, allowing the implementation of N-body initial conditions.

In Table 1 we summarize the characteristics of the initial galaxy pairs. 
Column (1) gives the run
identification code, where the number indicates the mass ratio, the lower-case letter denotes 
the impact parameter (h defines the head-on impact), and the capital letter indicates 
the energy of the orbit ({\it P} for parabolic, and {\it Z} for zero energy at infinity,
or hyperbolic orbit).
Hereafter, we will identify the encounter models with this code.
Column (2) gives the initial mass ratio of the colliding galaxies, 
column (3) gives their initial separation, 
column (4) gives their initial relative velocity,
and columns (5) and (6) give respectively the impact
parameter, and the orbital energy of the initial setup.

In Table 1, and later in describing the model parameters, we use the model units unless otherwise
noted.
We adopt non-dimensional units with with Newton's constant of gravity G=1. 
In each run, the theoretical half-mass radius of the Jaffe model, $r_{\rm J}$, and the total mass of the less
massive galaxy, $M_1$, are also set to 1. The models may be compared with real galaxies using the 
following scaling:

\begin{equation}
	        [M] = M_{\rm J} = 4\times10^{11} \; {\rm M_{\odot}}, \\
\end{equation}
\begin{equation}
		[R] = r_{\rm J} = 10 \;{\rm Kpc}  ,\\
\end{equation}
\begin{equation}
		[t] =  2.4\times10^7 \;{\rm yr}. \\
\end{equation}

By adopting these units, the velocity unit is:

\begin{equation}
		[v] =  414\; {\rm km s^{-1}}. \\
\end{equation}

Following Jaffe (1983) notation, the mass inside a radius r is defined as:

\begin{equation}
M(r)=\frac{r}{r+r_{\rm J}}  M\;  \\
\end{equation} 

In run $10hP$ the galaxy model with mass $M_2$ is a scaled up version of the 
model with mass $M_1$ with ten times more particles, constructed following the scaling 
relation between mass and radius given by Fish (1964):

\begin{equation}
\frac{M_1}{R^2_1} = \frac{M_2}{R^2_2} = K,
\end{equation}
where K is a constant. Following Jaffe's definition, and Fish's relation,
the theoretical half-mass radius for the more massive galaxy is 3.162. 

In order to limit the calculation time, and yet preserving the physical significance of the
results, we modify Jaffe's models in such a way to obtain {\it working} galaxy models with finite radii.
We impose a cut-off radius to all galaxy models, with $R_1=10$ for the less massive galaxies.
Such radius includes only 91$\%$ of the mass in the
theoretical Jaffe's model, thus we need to re-scale the half-mass radius as to keep
$M_1=1$. The re-scaled 
half-mass radius for the less massive galaxy is then equal to 0.82.

Model $1hP$  is an equal mass encounter between two galaxies on a parabolic orbit,
where $M_1=M_2=1$. 
The centers of the two galaxies are placed at an initial distance of $4~ R_1$. 
Model $10hP$ is an encounter between two galaxies with a mass ratio 10 on a radial 
parabolic encounter, initially placed at a distance $3~ R_1 +R_2$. $R_1$ and $R_2$ are the
galaxy radii of the working models. Model $1hZ$ is an encounter between two galaxies
with $M_1=M_2=1$,
placed initially on a mildly-hyperbolic orbit.

The choices of initial conditions for the galaxies are
 adequate to represent the observed mergers that occur
in elliptical galaxies. The galaxy mass ratios chosen for the 
simulations,
$M_2/M_1$=1 and 10, are the extremes of the observed merging galaxy 
sample by van Dokkum (2005).

\subsection{Stellar components}

We populate our model galaxies with model stars that represent stellar masses in the 
$m=0.85-8$ \ms~ mass range
\footnote{Hereafter, m refers to the stellar mass, to distinguish it from M, 
the galaxy mass},
whose progeny includes RGB, AGB, and PNe.
Since the main scope of our modeling is to study the stellar population that produce
PMS stars, we neglect all stars outside this mass range.
In order to model these stars we use test particles whose masses are proportional to 
the stellar masses they represent.
We assume the validity of Salpeter's (1955) initial mass function 
(IMF)
$\Psi(m)\propto m^{-2.35}$, and consider three 
representative
mass bins respectively for the progenitors of massive (3--8 \ms), intermediate
(1.4--3 \ms), 
and low-mass (0.85--1.4 \ms) PMS stars. 
The population of each mass bin corresponds to the 
integration of the Salpeter's mass function in that bin, scaled
to the entire population considered, and ignoring stars 
outside the mass range. We then calculate the mass fraction for each bin as:

\begin{equation}
\Phi_{\rm bin}= A^{-1} \int_{m_{\rm min}}^{m_{\rm max}} m ~\Psi(m) ~dm, 
\end{equation}
where m$_{\rm min}$ and m$_{\rm max}$ are the limits of the mass bin considered, and

\begin{equation}
A=\int _{0.85} ^{8.0} \Psi(m)~ dm = 0.8777
\end{equation}
is the normalization of Salpeter's law in the entire mass range.

To characterize the mass bins in the N-body simulation we use a representative mass for
each bin (1, 2, and 6 \ms). The model stars in the model galaxy with $M=1$
have such masses that, after accounting for the Salpeter's IMF, 
the total galaxy mass is equal to unity. 

In Table 2 we summarize the characteristics of the stellar population in the $M=1$ galaxies. 
Columns (1) gives the mass bin in solar masses, column (2) gives the representative mass
in that bin, colum (3) gives the mass fraction in
that mass bin, 
column (4) gives the number of particles in the bin, 
and column (5) the (dimensionless) stellar mass for each particle in that bin.
Note that the mass of each particle is different from the mass of the star it represents, 
but the ratio between particle masses in different bins is the same than the ratio of the 
representative masses in that bin. This is a good representation, since we 
are interested in relative results for the mass bins. 
In runs $1hP$ and $1hZ$ both 
galaxies have number of particles per mass bin as in Table 2. 
The least massive galaxy of run $10hP$ also has stellar population as in table 2, 
while the more massive galaxy has an initial set up with ten 
times more particles in each mass bin, but with the same mass per particle.

The initial distribution of the populations of particles with different mass
is such that each stellar population follows the same Jaffe law. In this way, there are 
fewer particles from the high mass bin at any given radius. We use the distribution
function (DF) as in Gonz{\' a}lez-Garc{\'{\i}}a \& van Albada (2005a), extending it 
for each mass bin.

\subsection{Integration method and stability}

We have used the parallel tree-code GADGET-1 (Springel et~al. 2001) on the Beowulf Cluster 
at the IAC, where a typical run on 16 CPUs takes of the order of $1.5 \times 10^5$ seconds. 
Gravitational Plummer Softening (see e.g. Binney \& Tremaine 1987, Eq. 2-194)
was set to $\sim~1/10$ of the half-mass radius of the less massive galaxy, with softening parameter
$\varepsilon = 0.075$. The tolerance parameter (see Barnes \& Hut, 1986) 
was set to $\theta=0.8$. Quadrupole terms were 
included in the force calculation.  GADGET-1 uses a variable time step ($t_{\rm s}$), and we choose
$t_{\rm s} \propto 1/|a|^{0.5}$, where $a$ is the particle acceleration.
We set the maximum time step to 1/100 of the half-mass crossing time,
and the minimum time step to zero.

We have checked the stability of our input initial model for $33$ crossing times. 
In Figure 1 we show the results from a test run that has been used to check the stability of our models. 
Therein, the evolution of the mass inside different radii has been followed at different mass-fraction levels,
showing stability in a large range of $t/\tau_{\rm cr}$, where $\tau_{\rm cr}$ is the half-mass crossing time. 
The test shows that the system relaxes for about 4 crossing times and remains stable thereafter. 
This initial relaxation is probably due to the presence of the particle softening in the code.

Models were evolved for at least $10$ dynamical crossing times of the merged system after 
merging, 
to allow the system to relax (reach virialization). Conservation of energy is sound in 
all the runs, with variations lower that 0.5$\%$ of the total energy.

We performed an additional run 
of a model identical to $1hP$, except with twice as many particles in each mass bin. Our merging
simulation on this additional model results in an intergalactic mass fraction of
5.22$\%$, 5.24$\%$, and 5.34$\%$ respectively for particles in 
the first, second, and third mass bin respectively, 
proving that our results are stable against mass resolution.

\begin{figure}

\plotone{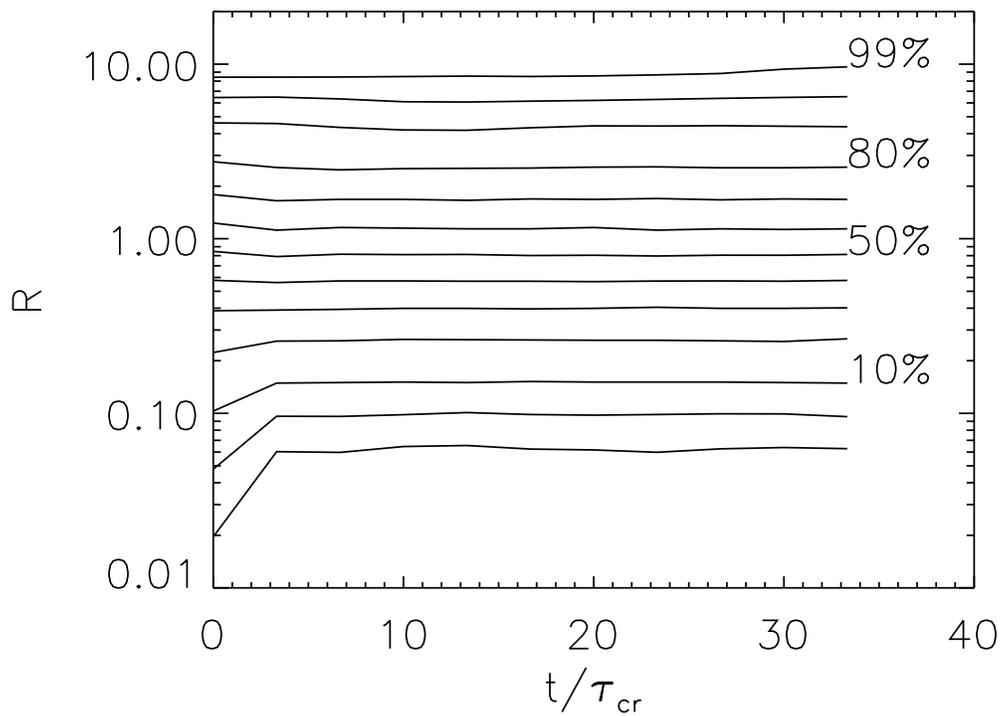}

\caption{A test run was made to check the stability of our initial models. 
The plot shows the evolution of the mass inside different radii, the top 
line gives the $99\%$~ 
mass radius, while the third line from the bottom gives the $10\%$~ mass radius.}

\end{figure}

\section{Results}

The three models of dry merging presented in this paper have different initial conditions both
in the mass content of the parent galaxies and in their relative velocities, as described in
$\S$2.
As model $1hP$ is let evolve, the system passes through pericenter for the first time
at t=85, when an
exchange of orbital energy into internal energy occurs. The particles having initial
binding energy close to zero gain enough energy at this stage
to be expelled from the system and become unbound particles. 
The two galaxies finally met again 30 time units after the first
encounter. 
This is the time when the actual merging occurs. New particles are able to become
unbound at this time, as a consequence of new particle encounters and exchange of energy.
The merging time is
$2.8\times 10^9$ yr after the initial placement in orbit, with the unit convention given in $\S$2.1.
A very similar situation occurs in model $10hP$, where the two unequal galaxies {\it meet}
at pericenter after t=70 since the initial placement in orbit,
then meet again 131 time units after the first encounter, 
with merging time of $4.82 \times 10 ^9$ yr. The galaxies of model $1hZ$,
initially in
mildly-hyperbolic orbits, meet for the first time at t=34.5 and after 65.5 time units the second and final
encounter occurs, ending in the merging episode. The merging time for $1hZ$ is then $2.4 \times 10^9$ yr.
All runs are stopped after ten merging crossing times, after virialization has been reached.

In Table 3 we give the characteristics of the stellar populations of our models after the 
merging has occurred. We give the run code (column 1), the mass bin (column 2), 
and the fraction of resulting unbound mass (column 3). We also give the unbound starlight
fractions in columns (4) and (5), as described in the Discussion.

\begin{figure}

\plotone{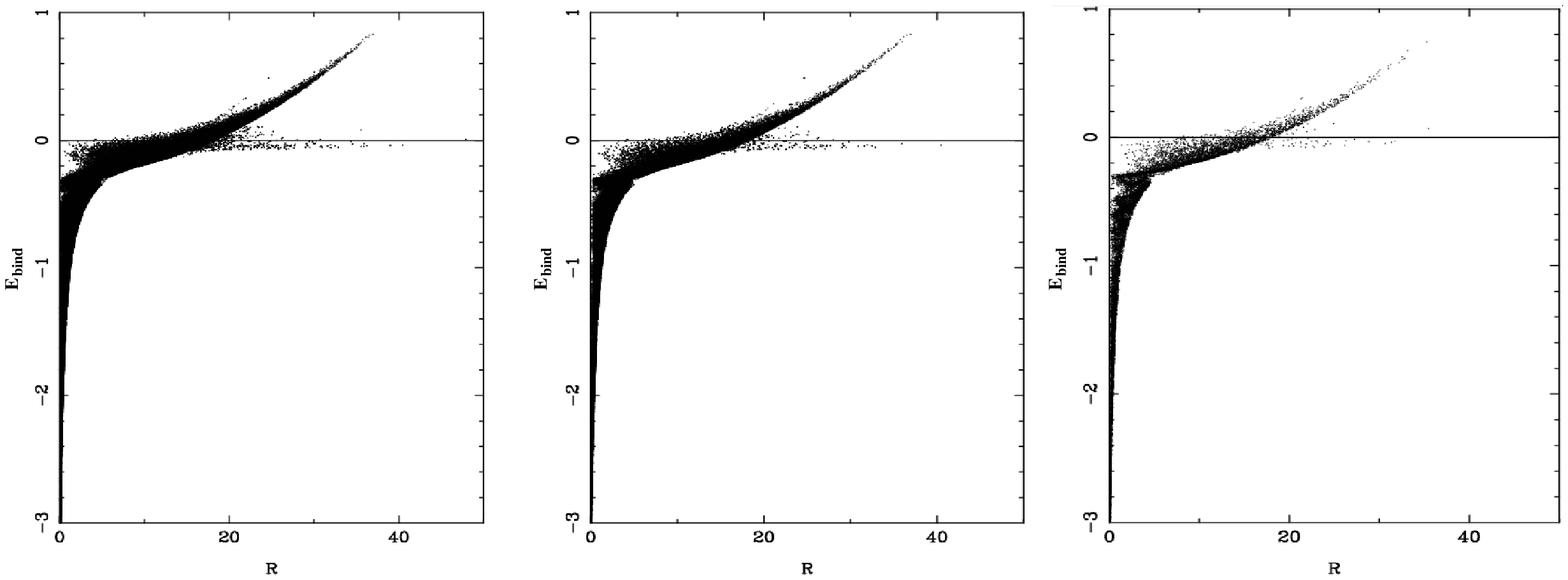}

\caption{Binding energy versus distance (from the merger's center)
of the model stars in the 1hP model. The panels, left to right, 
represent the situation in the low, intermediate, 
and high-mass bins.}

\end{figure}

\begin{figure}

\plotone{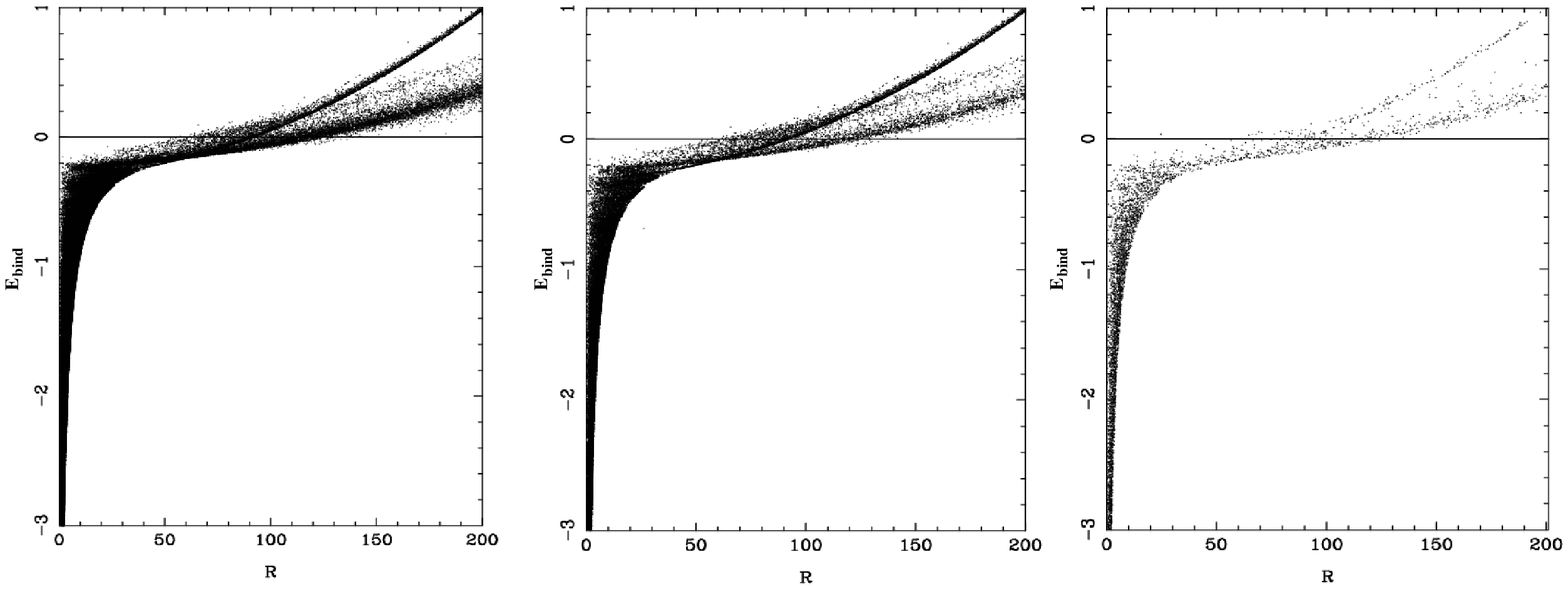}

\caption{Same as in Figure 2, but for the 10hP model.}

\end{figure}

\begin{figure}

\plotone{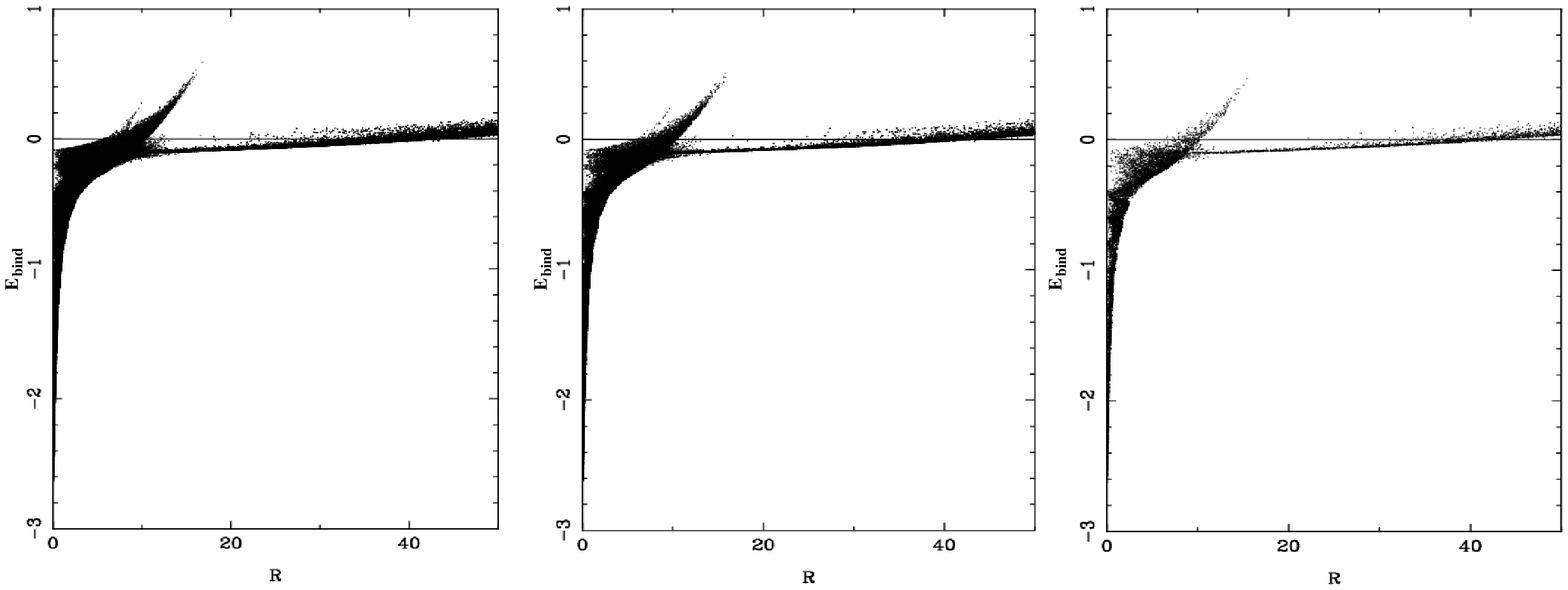}

\caption{Same as in Figure 2, but for the 1hZ model.}

\end{figure}

By inspecting the data of Table 3 we find several interesting results from
our merging simulations. First, we see that for all runs the percentage of
particles that become unbound does not depend on the mass bin. This result is expected,
since the 
particle expulsion is due to gravitational acceleration, and indicates that
if we were able to observe all IG stars with the same probability, we should 
recover the IMF of the original galaxies. Second, the final fraction of unbound mass is different for the three merging models:
in $1hP$ the fraction of unbound to total mass is 
$M_{\rm IG}/M_{\rm tot}=(5.48 \pm 0.1)\times
10^{-2}$, which means that $\sim 95 \%$  of the total initial mass still remains in the
merger ($M_{\rm m}=1.89$ with $M_{\rm tot}=2$). In run $10hP$ the fraction of unbound to total 
mass is almost twice as high, with
$M_{\rm IG}/M_{\rm tot}=(9.97 \pm 0.4)\times 10^{-2}$, corresponding to a merger
with 90$\%$ of the initial mass ($M_{\rm m}= 9.9$ for $M_{\rm tot}=11$).
Naturally, the larger fraction of unbound mass in model $10hP$ compared to $1hP$
could be imputed both to the initial mass ratio, but also to the high kinetic
energy due to the higher initial relative velocity in model $10hP$. The third model, $1hZ$, has the highest efficiency in producing unbound stars. 
The IG stellar mass in this case 
represent 21$\%$ of the total initial mass, and the merger retains only 79$\%$ of the 
initial mass. In this case, the merger has $M_{\rm m}=1.58$ from an initial
total mass $M_{\rm tot}=2$.

In Figures 2, 3, and 4  we show the location of the model stars
in the
radius -- binding energy plane
for each of the mass bins, for the $1hP$, $10hP$, and $1hZ$ models respectively. 
Both R and $E_{\rm bind}$ in the plot
are dimensionless variables, R being the distance from the merger's center.
Note that the plots do not include the complete range where particles are found, 
rather the range where most particles are found.
For example, out to a radius R=50 we find 90$\%$ of the $1hZ$ particles, and
up to 99$\%$ of the $1hP$ particles. These figures give a clear idea 
on the nature of the merger, and on the distribution of bound and unbound 
particles for each mass bin in each model considered. 
Unbound particles are found out to a distance of R$_{\rm max}\approx$160, 600, and
200 respectively for models $1hP$, $10hP$, and $1hZ$. 
We do find a mild anticorrelation between the particle mass and $R_{\it max}$.
By running a model similar to $1hP$ but
with twice as many particles per mass bin, we proved that the anticorrelation
is due to the 
relative populations of the bins, and it does not have physical significance.

In Figure 5, 6, and 7 we plot the surface density calculated within annuli
of increasingly larger radii from the
merger's center in models $1hP$, $10hP$, and $1hZ$ respectively. In these figures
we plot the logarithm of the surface density $\mu$~
against $R^{1/4}$, where $R$ is the distance from the merger's center. The three lines represent the 
different mass bins of stellar populations. Note that these plots are non cumulative, and that the
surface density have been evaluated in concentric annuli equally spaced in log$R$.
This representation is useful to show where the galaxy merger profile dominates,
and where the de Vaucouleurs (1959) profiles start to be perturbed by the unbound particles.
By comparing the $1hP$ (Fig.~5) and $1hZ$ (Fig.~7) models, we see that the surface 
density profile is similarly
perturbed, but the IG component outside the $R^{1/4}$=2 annulus is 
much more important in the $1hZ$ than in the $1hP$ model. The
de Vaucouleurs' slope is affected by the IG particles outmost of $R^{1/4}\sim 2.5$ in the
$10hP$ model (Fig.~6).
In all models, test particles from the different mass bins contribute to the
merger and the IG mass in very similar fashion.

\begin{figure}

\plotone{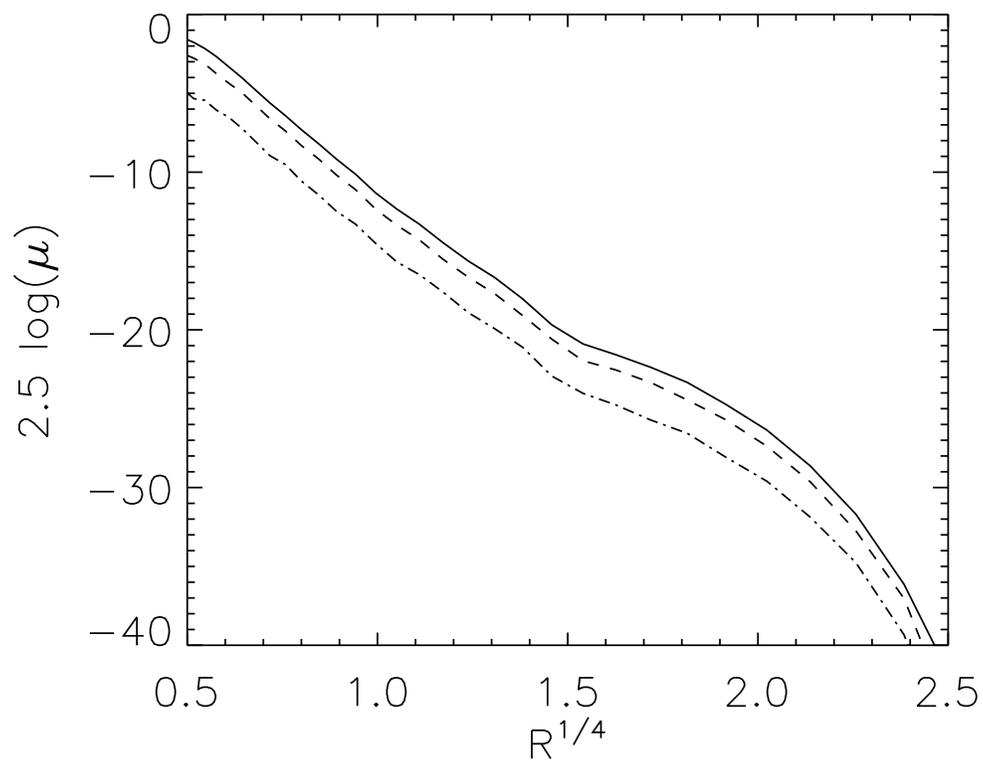}

\caption{Surface density versus $R^{1/4}$ in model $1hP$. 
The three lines represent the mass bins, where the solid line
represents the low mass bin, the dashed line the intermediate mass bin, and the 
dash-dot line the high mass bin. }

\end{figure}

\begin{figure}

\plotone{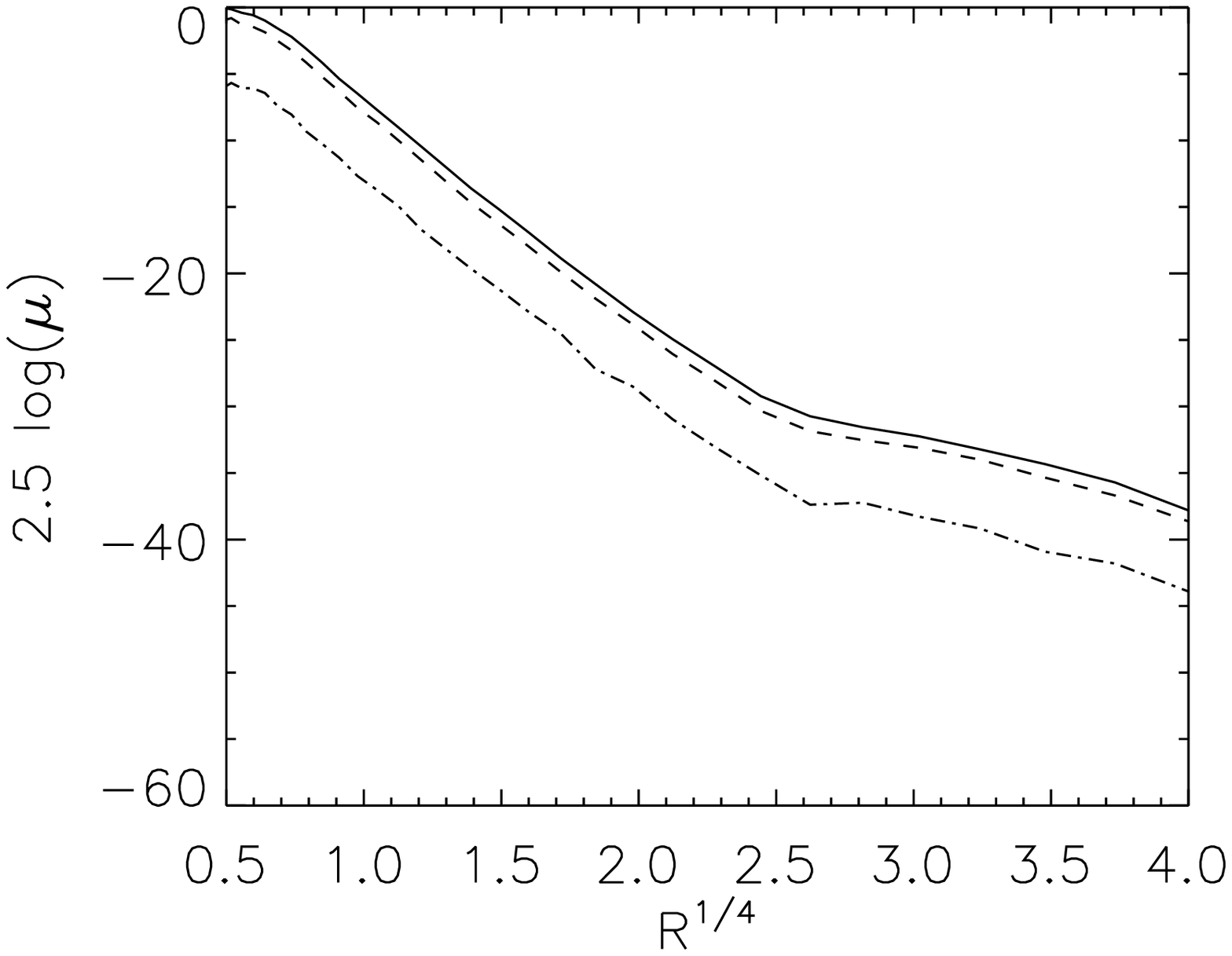}

\caption{Same as in Figure 5, for model 10hP.}

\end{figure}

\begin{figure}

\plotone{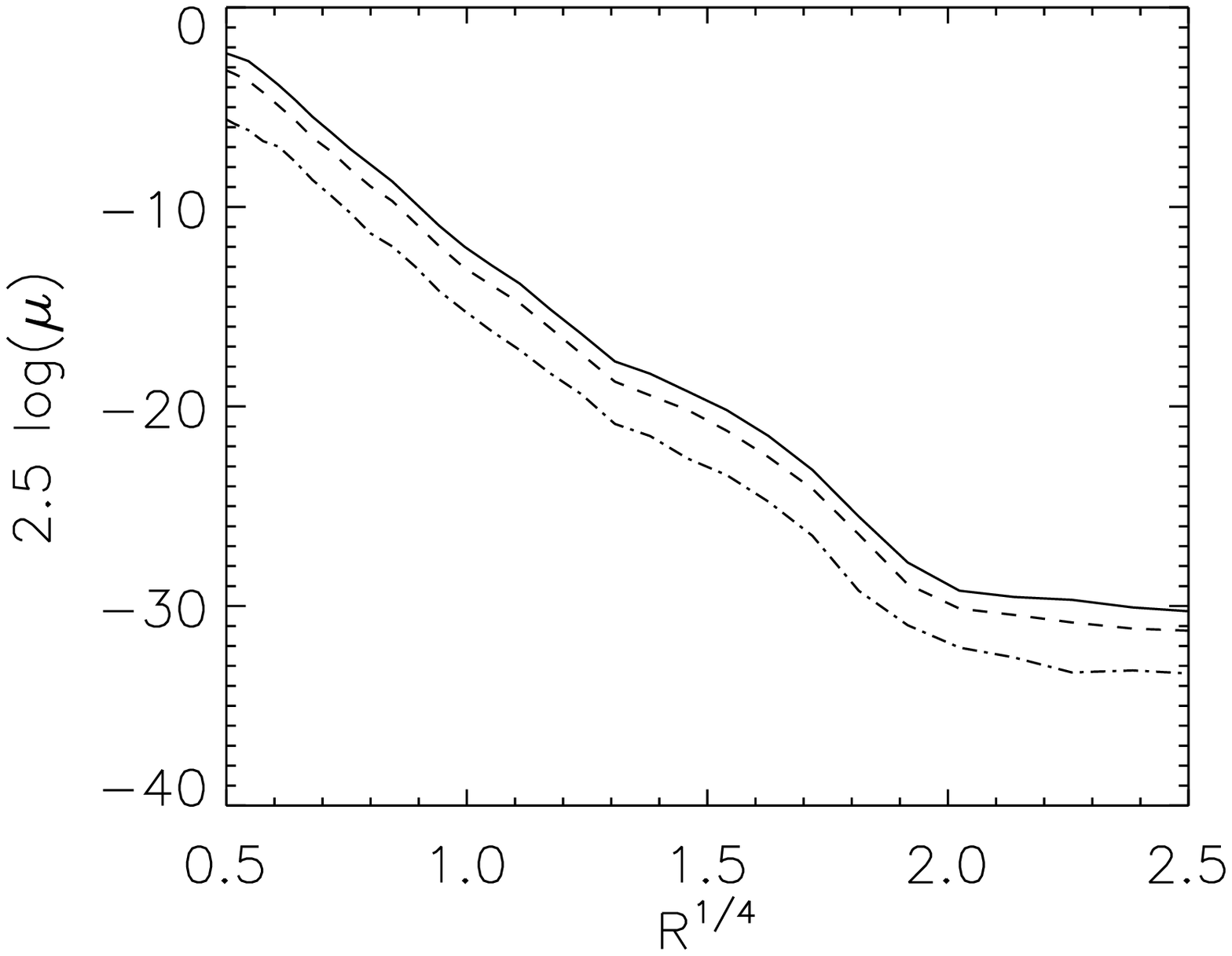}

\caption{Same as in Figure 5, for model 1hZ.}

\end{figure}

\section{Discussion}

Elliptical galaxy encounters were modeled
with N-body simulation, and they produce mergers in a fraction of the Hubble 
time. During the process
they expel a fraction of their stellar content.
The model stars expelled continue to evolve in the IG medium, eventually
going through the red giant, AGB, and post-AGB phases. 
The fraction of unbound mass resulting through this process
depends on the properties of the encounter. 
By comparing the two models with the initial parabolic orbits and different
mass ratios we conclude that the model with the higher mass ratio
produce a larger {\it fraction} of unbound mass with respect to the total
(as well as the merger's) mass. 
By comparing the $M_2/M_1=1$ runs we see that the initial hyperbolic orbit 
results in a unbound mass fraction that is almost four
times that of the parabolic orbit encounter.

Since the elliptical mergers belong to the fundamental plane of elliptical 
galaxies (Gonz{\' a}lez-Garc{\'{\i}}a \& van Albada 2003), in the same way than 
the two original merging galaxies, we can determine the fraction of
unbound-to-total starlight for all our models, simply by assuming 
that both the initial galaxies and the final mergers are homologous galaxies. 
If we assume that every galaxy of the merging pairs and all
mergers obey the mass-luminosity relation 
$M/L \propto M^{0.2}$ (Jorgensen et~al. 1996)
we can calculate 
$L_{\rm IG}/L_{\rm tot}$, the
ratio of the IG starlight over total starlight. 
In dry merging encounters the relation linking the total to individual galaxy luminosity
can be written as
$L_{\rm IG}=(L_1+L_2-L_{\rm m})$.
By using the mass fractions of unbound mass from Table 3 
we found $[L_{\rm IG}/L_{\rm tot}]_{1hP}$=17$\%$,  
$[L_{\rm IG}/L_{\rm tot}]_{10hP}$=14$\%$, and 
$[L_{\rm IG}/L_{\rm tot}]_{1hZ}$=28$\%$. 
The fractions of the unbound to merger's starlight are respectively
$[L_{\rm IG}/L_{\rm m}]_{1hP}=20\%$,
$[L_{\rm IG}/L_{\rm m}]_{10hP}=17\%$, and $[L_{\rm IG}/L_{\rm m}]_{1hZ}=38\%$
(see also Table 3).

By recalling that our model galaxies are populated by low- and intermediate-mass
stars, their $L_{\rm IG}/L_{\rm m}$ fractions may be compared with 
the observed PMS populations.
The IG starlight ratios that we recover in our simulations
are in broad agreement with the observations of several IC and 
intra-group stellar populations. Following, we examine a few particular cases.

The most studied case of IG stellar population is that of PNe in Virgo. 
The mechanism proposed in this paper is probably more likely to occur as a possible
origin of the IC stellar population, or parts thereof, in the cluster periphery 
rather than at the center of a cluster such as Virgo. 
Arnaboldi et~al. (2004) have studied the velocity dispersion of a sample of Virgo IC 
PNe and found that several fields have dispersion velocities much lower (247 km s$^{-1}$)
than the canonical Virgo dispersion of 800 [km s$^{-1}$], obviously a consequence of the
fact that the cluster is young and highly nonuniform.  Feldmeier
et~al. (2004b) found that $\sim 16 \%$ of the starlight in the Virgo cluster
is in the IG medium, independent on the location within Virgo,
and this number is encompassed by our results.

Nonetheless, the observed ratio of unbound to total starlight in the Virgo cluster derived from
PNe must be used {\it cum grano salis}. To derive this ratio from observations one must
evaluate the theoretical luminosity-specific PN density, $\alpha_{\rm PN}=B~t_{\rm PN}$ 
(Renzini \& Buzzoni 1986), which is based in the {\it fuel consumption
theorem} for PMS stars. The question is 
whether this theorem is adequate to describe a nebular population, where $t_{\rm PN}$, the
PN lifetime, is not a stellar evolutionary time, determined by fuel consumption, but rather
a time-scale depending on the hydrodynamic evolution and the photo-ionization on the
nebulae.  While $t_{\rm PN}=25,000 ~yr$ is typically adopted, the correct lifetime for PNe to 
be observable at high luminosity is probably much
lower, as the hydrodynamic models by Villaver \& Stanghellini (2005) have shown.
Furthermore, $t_{\rm PN}$ has not been parametrized 
for the mass and chemistry of the progenitor stars. Feldmeier et~al. (2004b)
advise to use $\alpha_{\rm stars}$ from other stellar 
sources, such as Virgo red giants (Durrell et~al. 2002). 

The red giant population of the Virgo IC, first observed by Ferguson et~al. (1998), 
accounts for approximately 10$\%$ of the cluster (evolved) stellar
mass. Ferguson et~al. (1998) indicated that the IG population
is likely to originate from elliptical and S0 galaxies, for their higher frequency in the cluster
and their older stellar populations.                                                                           
Intergalactic starlight has been observed in Fornax both in the form of PNe and other 
stars. Given the difficulty of
PNe lifetime scaling, we prefer to use the results from IG novae observations
(Neill et~al. 2005),
indicating that $\sim$16 to 41 $\%$ of the starlight in the Fornax cluster comes from IG stars. 
Both the Virgo and Fornax PMS IG stars might have dry merging origin, and these percentages
are clearly in the range of our results (see Table 3, column 4). Group IG populations are typically evaluated to be 10 times lower than their cluster counterparts,
although spectroscopic confirmation of, for example, the M81 PNe has not yet been published
(Feldmeier et~al. 2004a) and final counts are not available. 

The mere existence of low-mass PMS stars in a given stellar population requires a very old
stellar population. Following Maraston's (1998) prescription, the turnoff mass in the
first star mass bin correspond to ages in the $\sim$4.8--22 Gyr range. If, for sake of reasoning, we
assume that the progenitor galaxies were just formed at the time they were put in their
relative orbits, none of their low-mass (0.85-1.4 \ms) stars would have reached the turnoff 
by the time the merger is completely formed. More realistically, the merging galaxy pairs contain 
aging stellar populations when their first encounter occurs. 
Accordingly to van Dokkum and collaborators (van Dokkum et~al. 1999, 
van Dokkum 2005), dry merging could have been important at intermediate redshifts
when clusters were still assembling. 
Therefore, part of the IC stellar population may have been expelled 
at those cosmic times.

Zibetti et~al.~(2005) found that the intracluster light at large cluster
radii is largely dominated by surface brightness excess around 
galaxies. From our Figure 5 we find that the IC contribution to the profile
is above the $R^{1/4}$ law from $R^{1/4}\approx1.5$ to 2. This implies that the IC stars 
contribute the most to the surface density out to $\approx$100 Kpc from the
merger edge of model $1hP$ (or, with similar reasoning, about 60 Kpc from the 
merger edge in model $1hZ$) with the usual unit conversion, consistent with 
what observed by Zibetti et al. Naturally this analysis is sensitive to the choice of units
that we use to compare the models with real galaxies, as described in $\S$2, 
but it is worth showing that our equal-mass encounter models are at least broadly consistent
with the observations. 

In this paper we have compared our models with the observed IG populations.
In other words, we have implicitly assumed that all IG stars derive from dry merging, 
and that all galaxies in a given observed cluster or group has
gone through at least one merging process. In this extreme assumption the models 
correctly predict the observed IG starlight.
Naturally, the case might well be that dry merging occurs only in a fraction
of cases, and that other mechanisms are at work in explaining the observed unbound starlight
in galaxy associations. While the 
dry merging scenario is certainly helpful to account for a fraction of the observed IG light,
it may not work well near cluster centers and in high velocity environments, where other
mechanisms
such as mass stripping due to hyperbolic encounters, or disk galaxy merging,
might be more efficient. 

In the future, we plan to model similar encounters as those presented here, but
including dark halos in elliptical galaxies. Furthermore,
other phenomena related to dry merging, and a larger range of initial conditions, will be considered in future studies.

\acknowledgments

We warmly thank Dr. Luca Ciotti for important comments on the 
manuscript and
for scientific discussions, and Drs. Mark Dickinson and John Feldmeier for 
scientific discussion and bibliographic indications. L. S. 
thanks the European Southern Observatory and the IAC for their hospitality 
during the summers of 2004 and 2005, when this work was conceived and
completed.

\clearpage
\begin{deluxetable}{lccccc}
\tablewidth{9truecm}
\tablecaption{Model input, galaxy pairs parameters}
\tablehead{
\colhead {Run} & 
\colhead{M$_2$/M$_1$}& 
\colhead{$r_i$}&
\colhead{$v_i$}&
\colhead{b} & 
\colhead {E$_{\rm orb}$} \\
}
\startdata
$1hP$ & 1 & 40 & 0.316& 0 & 0 \\
$10hP$ & 10 & 61.62& 0.604&0 & 0\\
$1hZ$ & 1 & 40 & 1.048& 0 & 0.250 \\
\enddata
\end{deluxetable}

\clearpage

\begin{deluxetable}{lrrrr}
\tablewidth{10truecm}
\tablecaption{Model input, stellar parameters}
\tablehead{
\colhead{bin [\ms]} &
\colhead{M [\ms]}&
\colhead{$\Phi_{bin}$}& 
\colhead{N$_{particles}$} &
\colhead{m$_{particle}$ }\\
}
\startdata
0.85 -- 1.4 & 1	& 0.5151& 156060 &  $3.3\times10^{-6}$\\
1.4 -- 3.0 &  2	& 0.3443& 52121&	$6.6\times10^{-6}$\\
3.0 -- 8.0 &   6	& 0.1406& 7050&	$2.0\times10^{-5}$\\
\hline
total& &1& 215231& 1\\ 

\enddata
\end{deluxetable}

\clearpage
\begin{deluxetable}{lcrrcc}
\tablewidth{10truecm}
\tablecaption{Model results}
\tablehead{
\colhead{Run}&
\colhead{bin }&
\colhead{M$_{\rm IG}$} &
\colhead{$L_{\rm IG}/L_{\rm tot}$}&
\colhead{$L_{\rm IG}/L_{\rm m}$ }
\\
\colhead{}& \colhead{[\ms]}& \colhead{$\%$}& \colhead{$\%$}&\colhead{$\%$}\\
}

\startdata
$1hP$&	0.85 -- 1.4&	  5.53 &&\\
$1hP$&	1.4 -- 3.0&	  5.55 &&\\
$1hP$&	3.0 -- 8.0&	   5.37&&\\
$1hP$&	total&	  5.48 & 17 &20\\

\hline

$10hP$&	0.85 -- 1.4&    9.87 &&\\
$10hP$&	1.4 -- 3.0&    10.20 &&\\
$10hP$&	3.0 -- 8.0&      9.57 && \\
$10hP$&	total& 9.97 & 14& 17\\

\hline
$1hZ$&	0.85 -- 1.4&    20.6&& \\
$1hZ$&	1.4 -- 3.0&    20.6 &&\\
$1hZ$&	3.0 -- 8.0&      20.9 &&\\
$1hZ$&	total& 		20.7& 28& 38\\

\enddata
\end{deluxetable}

\end{document}